\documentclass[prb,showpacs,superscriptaddress,amsmath,amssymb,twocolumn]{revtex4-1}
\usepackage{graphicx}
\usepackage{graphics}
\usepackage{dcolumn}
\usepackage{bm}
\usepackage{pstricks}
\usepackage{float}
\usepackage{amssymb}
  \usepackage{hyperref}
\usepackage[normalem]{ulem}

\definecolor{lineadecolor}{rgb}{0.35,0.5,0.6}
\definecolor{ddcol}{rgb}{0.8,0.1,0.1}
\definecolor{subsectioncolor}{rgb}{0.1,0.01,0.5}
\definecolor{celeste}{rgb}{0.8,0.87,0.99}

\newcommand{\ba}{\begin{eqnarray}}
\newcommand{\ea}{\end{eqnarray}}
\def\be{\begin{equation}}
\def\ee{\end{equation}}

\begin{document}

\title{Current jumps in flat band ladders with Dzyaloshinskii-Moriya interactions}

\author{ S.\ Acevedo}
\affiliation{IFLP - CONICET. Departamento de F\'isica, Facultad de Ciencias Exactas. Universidad Nacional de La Plata,C.C.\ 67, 1900 La Plata, Argentina.}

\author{ P.\ Pujol}
\affiliation{Laboratoire de Physique Th\'eorique-IRSAMC, CNRS and Universit\'e de Toulouse, UPS, Toulouse, F-31062, France}

\author{ C.A.\ Lamas}
\email{lamas@fisica.unlp.edu.ar}
\affiliation{IFLP - CONICET. Departamento de F\'isica, Facultad de Ciencias Exactas. Universidad Nacional de La Plata,C.C.\ 67, 1900 La Plata, Argentina.}

\begin{abstract}
Localized magnons states, due to flat bands in the spectrum, is an intensely studied phenomenon and can be found in many frustrated magnets 
of different spatial dimensionality. 
The presence of Dzyaloshinskii-Moriya (DM) interactions may change radically the behavior in such systems. In this context, we study a paradigmatic 
example of a  one-dimensional frustrated antiferromagnet, the sawtooth chain in the presence of DM interactions. 
Using both path integrals methods and numerical Density Matrix Renormalization Group, we revisit the physics of localized magnons and determine the consequences of the DM interaction on the ground state.  We have studied the spin current behavior, finding three different regimes. 
First, a Luttinger-liquid regime where the spin current shows a step behavior as a function of parameter $D$, at a low magnetic field. Increasing the magnetic field, 
the system is in the Meissner phase at the  $m=1/2$ plateau, where the spin current is proportional to the DM parameter.
Finally, further increasing the magnetic field and for finite $D$ there is  a small stiffness regime where the spin current shows, at fixed magnetization, a jump to large values at $D=0$, a phenomenon also due to the flat band.

\end{abstract}

\maketitle

\section{Introduction}

In low dimensional systems, geometrical frustration and quantum fluctuations may lead to unusual magnetic states. 
In two dimensions, Magnetic materials with Kagom\'e lattice structure have attracted much attention in the field of condensed matter physics due to
their exotic magnetic phenomena. Some realizations of the $S=1/2$ Kagom\'e lattice are the Herbertsmithite ZnCu$_3$(OH)$_6$Cl$_2$ , $\alpha$-vesignieite
BaCu$_3$V$_2$O$_8$(OH)$_2$,  and [NH$_4$]$_2$ [C$_7$H$_{14}$N][V$_7$O$_6$F$_18$]$_5$. 
An interesting phenomenon that is known for quite a few years, and among which the Kagome lattice provides an example, corresponds to frustrated systems
with a flat band in the energy spectrum.\cite{schulenburg2002macroscopic,acevedo2019magnon}. 
However, the presence of flat bands is not exclusive for two-dimensional systems. 
It is possible to find one-dimensional systems, where localized magnons excitations emerge due to frustration\cite{acevedo2019magnon}. 
Here we consider the presence of the Dzyaloshinskii-Moriya (DM) interaction, representing the antisymmetric version of the Heisenberg exchange
induced by the spin-orbit coupling, in frustrated one-dimensional systems. More precisely, we focus on the paradigmatic antiferromagnet in one dimension
known as the sawtooth chain, but we also explore the consequences of frustration in other similar systems.

In the absence of the Dzyaloshinskii-Moriya interactions, the Heisenberg model on the sawtooth chain presents two degenerate ground states at $M = 0$ 
and the elementary excitations were found to be ‘‘kink’’-‘‘antikink’’-type excitations.\cite{nakamura1996elementary,sen1996quantum}. 
Hao {\it et. al.} show that a weak DM interaction is sufficient to break the valence-bond order (VBO) and lead the system into a Luttinger liquid with
algebraic spin correlations\cite{hao2011destruction}. 
In the same chain, the Bose Hubbard model presents a solid phase which is unstable against doping\cite{huber2010bose}. 
The same model on different ladders geometries may present many states, including Meissner phases, vortex
fluids, vortex lattices and charge density waves.\cite{greschner2016symmetry,Vekua-current-reversal,Laflorenci-HC-bosons,white-bose,Orignac_2016}.

For the Heisenberg model, it is possible to construct exact eigenstates of independent localized one-magnon 
states that become the ground state previous to saturation in a family of one dimensional spin systems for which the sawtooth
chain is the simplest member\cite{schulenburg2002macroscopic,schmidt2002linear}.


 
\begin{figure}[t!]
\begin{center}
  \includegraphics[width=0.42\textwidth]{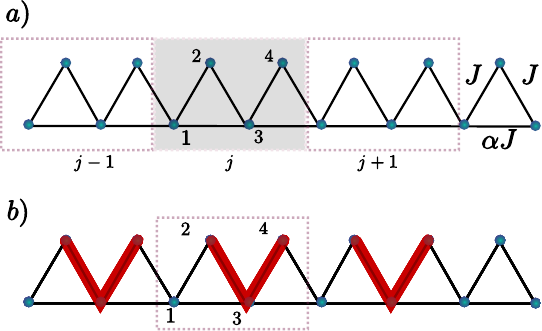}
\caption{Sketch of the Sawtooth chain. a) A four sites unit cell is used along the paper. b) Representation of the Magnon crystal ground
state at the critical coupling $\alpha_{c}$.  }
\label{fig:sawtooth}
\end{center}
\end{figure}

The sawtooth geometry is shown in Fig. \ref{fig:sawtooth}. This lattice presents a strong geometrical frustration modulated by the parameter $\alpha$ (see Fig. \ref{fig:sawtooth}-a).
We will show that DM interactions produce a nontrivial behavior in the spin current in the high magnetic field regime.

The paper is organized as follows: In section \ref{sec:magnon-jump} we present the Hamiltonian for the sawtooth chain and a summary of the exact results for  
the magnetization jump above the plateau in the presence of the localized magnons. In Section \ref{LEEM} we presents a coherent-state path integral
description on the sawtooth chain and we determine a condition to obtain a non-trivial delocalized mode trigering short range entanglement and an emergent 
spin imbalance, a phenomenon among which localized magnons is a particular case. 
In section \ref{sec:dmrg} we present Density Matrix Renormalization Group calculations of the magnetization process and the spin current corresponding
to different values of the DM coupling. Three different behaviors are identified  corresponding to low, medium and high magnetic field. 
In section \ref{sec:kagome-ladder} we apply the path integral results to other Kagom\'e-like spin ladders  and finally in section \ref{sec:conclutions} 
we present the conclusions.

\section{Localized Magnons and magnetization jumps.}
\label{sec:magnon-jump}

Let us consider the Heisenberg Hamiltonian on the Sawtooth chain given by

\begin{eqnarray}
\nonumber
    H & = & J  \sum_j \left( \boldsymbol{S}_{j,1}^{\Delta} \boldsymbol{S}_{j,2}^{\Delta}
    +
    \boldsymbol{S}_{j,2}^{\Delta} \boldsymbol{S}_{j,3}^{\Delta}
    +
    \boldsymbol{S}_{j,3}^{\Delta} \boldsymbol{S}_{j,4}^{\Delta}\right. \\
    \label{eq:sawtooth_hamiltonian}
    &+& \left. \alpha \big(
    \boldsymbol{S}_{j,1}^{\Delta} \boldsymbol{S}_{j,3}^{\Delta}
    +
    \boldsymbol{S}_{j,3}^{\Delta} \boldsymbol{S}_{j+1,1}^{\Delta}
    \big )
    +
    \boldsymbol{S}_{j,4}^{\Delta} \boldsymbol{S}_{j+1,1}^{\Delta}
    \right)\\\nonumber 
    &-&h\sum_{j,k}S_{j,k}^z    
\end{eqnarray}
with $j=1,...,L$ denoting the cell index, $\alpha$ regulates the frustration in the triangles and $k=1,...,4$ is the internal index in each unit cell as 
shown in Fig. \ref{fig:sawtooth}. 
Here exchange terms $\boldsymbol{S}_{i,l}^{\Delta} \boldsymbol{S}_{j,m}^{\Delta}$ denotes the spin interaction in the presence of anisotropy $\Delta$

\begin{equation}
 \boldsymbol{S}_{i,l}^{\Delta} \boldsymbol{S}_{j,m}^{\Delta}=\frac{1}{2}(S^{+}_{il} S^{-}_{jm}+S^{-}_{il} S^{+}_{jm})+\Delta S^{z}_{il} S^{z}_{jm}.
\end{equation}

In the absence of Dzyaloshinskii-Moriya interactions, the lowest magnon branch for the sawtooth chain becomes flat by tuning the couplings
at $\alpha=\alpha_{c}=1/\sqrt{2(1+\Delta)}$. At this point, a magnon in a unit cell is completely decoupled from the rest of the chain. 
The ground state corresponds to
a product state $|g.s.\rangle=\frac{1}{\mathcal{C}}\prod_{j}|\psi_{j}\rangle$ with local one-magnon states given by
$$|\psi_{j}\rangle=\sum_{k\in cell_{j}}\lambda_{k}\hat{S}_{k}^{-}|\uparrow\uparrow\uparrow\cdots\uparrow\uparrow\uparrow\rangle$$
where coefficients $\lambda_{k}$ 
are nonzero only for $k=2,3,4$ within the unit cell, as highlighted with red thick lines in Fig. \ref{fig:sawtooth}-b.

As schematized in Fig. \ref{fig:sawtooth}-b we can construct further local excitations and there will be no interaction between excitations 
as long as they are separated in space. In this way we obtain $n$-magnon excitations  whose energy is $n$ times the energy of one isolated magnon. This
multiple magnon state becomes the lowest magnon excitation\cite{schnack2001independent}. The analytical proof of this statement it is not easy, but the numerical evidence is clear.

In Ref. \onlinecite{schulenburg2002macroscopic} the magnetization process for the sawtooth chain with a flat band, in the absence of DM interactions, 
is described together with two other ladders with Kagom\'e-like structure (that we will introduce in \ref{sec:kagome-ladder} and can also have a flat band)
and the 2D Kagom\'e lattice. In section \ref{sec:dmrg}  we show the normalized magnetization as a function of the external magnetic field. In the absence of Dzyaloshinskii-Moriya 
interactions a 
macroscopic jump from $m=1/2$ to $m=1$  makes evident that the lowest excitation to the fully polarized states contains several magnons. This jump with $\delta m=1/2$ 
corresponds
to a configurations of magnons with a four spins unit cell like the one schematized in Fig. \ref{fig:sawtooth}-b.

The XXZ anisotropy in the model does not affect properties of the one-magnon dispersion, and then the degeneracy and
the associated magnetic jump is expected to be independent of $\Delta$. However, in order to construct a low energy theory, it is convenient to study the system
for $\Delta\neq 1$ to avoid classical collinear configurations that disfavor the  path integral formulation  in terms of spin coherent states. For this reason,
in the following we will work with $\Delta = 1/2$ unless otherwise indicated.

\section{Low energy effective model}
\label{LEEM}

We study the system using the coherent-state path integral description due to Haldane\cite{Haldane1986} and Tanaka et. all\cite{TTH}. 
For an introductory and detailed approach to this description see references\onlinecite{TTH,auerbach2012interacting,PI1}.
In order to obtain an effective theory, first we identify the classical lowest energy configuration. 
To do this, we start from the isolated triangle formed by the $\boldsymbol{S}_1,\boldsymbol{S}_2,\boldsymbol{S}_3$ spins in a given unit cell (see Fig. \ref{fig:sawtooth}).
By symmetry, we take the $\boldsymbol{S}_1$ and $ \boldsymbol{S}_3$ polar angles to be equal, i.e. 
$\theta_1=\theta_3\equiv \theta_B$, and $\phi_2-\phi_1=\phi_2-\phi_3\equiv\phi$ for the azimuthal angles. For the complete chain we take $\theta_2=\theta_4\equiv \theta_A$ in each cell, and a unique azimuthal angle as well.

Choosing the anisotropy paremater for example to $\Delta=1/2$, the classical ground state for a given applied magnetic field consists in a canted configuration,
with $\phi= \pi$, and $h$-dependent $\theta_{A,B}(h)$ as usual\cite{TTH,PI1,PI2}. We write the spin operators in terms of the polar and azimuthal angles as $O(3)$ vectors with length
$S$ (being $S$ the spin quantum number), by 

$$\boldsymbol{S}_{jl}=S(\sin\theta_{jl}\cos\phi_{jl},\sin\theta_{jl}\sin\phi_{jl},\cos\theta_{jl})$$
and we parametrize the fluctuations around the classical configuration as
\begin{equation}
\phi_{jl}\rightarrow \phi_{0l} + \phi_{l}(x_{j})\hspace{1cm} \\  \hspace{1cm} \theta_{jl}\rightarrow \theta_{0l} + \delta\theta_{l}(x_{j}).
\label{eq:fluctuations}
\end{equation}
The conjugate variables used to construct the effective field theory are   $\phi_l(x_j)$ and
\begin{equation} 
  a \Pi_l(x_j) = -S \big (  \delta \theta_l (x_j) \sin{\theta_{0l}} + \frac{1}{2} ( \delta \theta_{l}(x_j))^2  \cos{\theta_{0l}}\big )
\end{equation}
where $a$ is the lattice spacing.
The spin operators are then written as
\begin{equation}
 S_{j l}^z= S \cos{\theta_{0 l}} + a \Pi_{l}(x_j)
\end{equation}
\begin{equation}
    \begin{split}
        S_{j l}^\pm=S
        (-1)^l  e^{\pm i \phi_{l}(x_j)}
         & \bigg( \sin{\theta_{0 l} - \frac{a \Pi_{l}(x_j)}{S \tan{\theta_{0l}}}} \\
        & - \frac{1}{2} 
        \frac{1}{1-\cos^2{\theta_{0 l}}}
        \frac{a^2 \Pi_{l}^2(x_j)}{S^2 \sin{\theta_{0 l }}}
        \bigg)
    \end{split}
    \label{eq:spintofieldlowE}
\end{equation}
The theory is written up to quadratic terms in fluctuations. First order terms vanish because fluctuations are added upon the
classical ground state, and constant terms are drop.

The total action of the system  $ \mathcal{S}=\mathcal{S}_{cl}+\mathcal{S}_{BP} $ is  conected to the partition function of 
the system via 
a path integral over all possible spin trayectories in imaginary time $\tau$.
 $\mathcal{S}_{cl}=\int d\tau H (\tau)$
is the  classical action 
and $\mathcal{S}_{BP}$ is the Berry Phase term, a geometrical term that emerges due to the overcomplete nature of the coherent state 
basis, and depends on the spins trayectories, and not on their explicit time dependence\cite{auerbach2012interacting}. In the continuum description, 
the later is simply 
\begin{equation}
    \mathcal{S}_{BP}= i \int dx d\tau  \sum_{l=1}^4  \bigg \{(\partial_\tau\phi_l) \frac{S-m_{l}}{a}
    -   (\partial_\tau \phi_l) \Pi_l
    \bigg\}
\end{equation}
while the former is itself divided in kinetic and mass terms, i.e. $\mathcal{S}_{cl}=\mathcal{S}_K + \mathcal{S}_M$, where
\begin{equation}
 \begin{split}
  \mathcal{S}_{M}=   J \iint \frac{dx}{a} d\tau
  \bigg \{
  \frac{S^2}{2}  \vec{\phi}^t M_\phi \vec{\phi}  + 
  \frac{a^2}{2} \vec{\Pi}^t M_\Pi \vec{\Pi} +
  \bigg \}
 \end{split}
\end{equation} 
and $ m_l = S \cos(\theta_{0l})$, $(\vec{\phi})_i=\phi_i$, $(\vec{\Pi})_i=\Pi_i$, and $M_\phi$ and $M_\Pi$ are symmetric matrices depending on $h$, $\theta_A$, $\theta_B$ and $\alpha$. 
$\mathcal{S}_K$ will be presented below for clarity. We perform a diagonalization of the symmetric matrix $M_\phi$ using a unitary transformation $P$, such that $\vec{\varphi}=P \vec{\phi}$, where
\begin{equation}
    P=
   \left(
\begin{array}{cccc}
 \frac{1}{2} & \frac{1}{2} & \frac{1}{2} & \frac{1}{2} \\
 0 & -\frac{1}{\sqrt{2}} & 0 & \frac{1}{\sqrt{2}} \\
 -\frac{1}{2} & \frac{1}{2} & -\frac{1}{2} & \frac{1}{2} \\
 -\frac{1}{\sqrt{2}} & 0 & \frac{1}{\sqrt{2}} & 0 \\
\end{array}
\right),
\label{eq:Pmatrixcombsaw}
\end{equation}
obtaining $    M'_\phi= P M_\phi P^t=\text{Diag} \left ( 0,m_2,m_3,m_4\right) $ 
where
\begin{equation}
\begin{split}
     m_2&=2 \gamma \\
   m_3&= 4 \gamma \\
    m_4&=2  (\gamma - 2 \alpha \beta ) 
    \label{eq:masses}
\end{split}
\end{equation}
with $\beta=\sin^2{\theta_B}$ and    $\gamma=\sin{\theta_A} \sin{\theta_B}$. 

Eqs. \eqref{eq:Pmatrixcombsaw} and \eqref{eq:masses} mean that the field $\varphi_1=\big (\phi_1+\phi_2+\phi_3 +\phi_4 \big)/2$ is gapless, as a 
consequence of the $U(1)$ symmetry of the system. The fields $\varphi_2=\big(\phi_4 -\phi_2\big)/\sqrt{2}$ and
$\varphi_3=\big(-\phi_1+\phi_2-\phi_3+\phi_4 \big)/2$  are both gapfull and at  the low energy limit frozen to a vanishing value. 
Finally, the combination $\phi_3 -\phi_1$ can be gapples if 
\begin{equation}
\alpha =\frac{\gamma}{2 \beta}
\end{equation}

 If $m_4$ vanish, the field $\varphi_4=\big(\phi_3 -\phi_1\big)/\sqrt{2}$ it's free to fluctuate, i.e. it gets delocalized, and it's conjugate field 
$\Omega_4=\big(\Pi_3 -\Pi_1\big)/\sqrt{2}$ is then localized. If the $\Omega_4$ mass term also vanish, it can be energetically favorable for it to get locked into a non-zero value. This is the mechanism that was
proposed in \onlinecite{plat_spintube}  for a phenomenon of a spontaneous spin imbalance and a factorization of the wave functions when a magnetization plateau is present. Indeed,  $\Omega_4 \neq 0$ means that there is a difference in the local magnetization between spins 1 and 3, breaking the lattice symmetry. 
This scenario is indeed corroborated by the DMRG analysis as we show in section \ref{sec:dmrg}.

It is interesting to look also at  the kinetic part or the action, containing the spatial derivatives of $\phi$, we have
 \begin{equation}
 \begin{split}
    \mathcal{S}_K=\frac{JS^2}{2} & \iint \frac{dx}{a} d\tau \bigg\{
  ( \gamma - \alpha \beta ) \big( a \partial_x \phi_1(x) \big )^2  +  \\ 
  &2\big( a \partial_x \phi_1 (x) \big )  [\gamma(\phi_1 - \phi_4)-\alpha \beta (\phi_1 - \phi_3)]
  \bigg \}
 \end{split}
 \end{equation}
 The action exhibits first derivative terms which are not conventional kinetics terms. In general, because of the vanishing of the $\varphi_2$, $\varphi_3$ and $\varphi_4$ fields discussed above, these terms vanish in the low energy limit of the effective action.
In the interesting case giving rise to the delocalization of the $\varphi_4$ field, then the coefficient in front of the first derivative terms vanish. Then, at low energy, $S_K$ is simply.

 \begin{equation}
  \mathcal{S}_K=\frac{JS^2}{2} \iint \frac{dx}{a} d\tau \bigg\{
  ( \gamma - \alpha \beta ) \big( a \partial_x \phi_1(x) \big )^2   \bigg \}
  \label{eq:SKinetic_final}
 \end{equation}
The only stiffness coefficient, $ \gamma - \alpha \beta = \frac{1}{2} m_4 + \alpha \beta$ its minimum when $m_4$ vanish. The phenomenon of a flat band can be interpreted here as a (renormalized) stiffness vanishing simultaneously to the $\varphi_4$ and $\Pi_4$ mass terms discussed above. As we see below,
this produce a dramatic effect in the presence of DM interactions.

As it is known a DM interaction produces spin currents in the system. But one question that naturally arises is what will be the effect of 
the flat band on the currents? To answer this question we determine numerically the spin currents by DMRG.

\section{DMRG results}
\label{sec:dmrg}

\begin{figure}[t!]
\includegraphics[width=1.0\columnwidth]{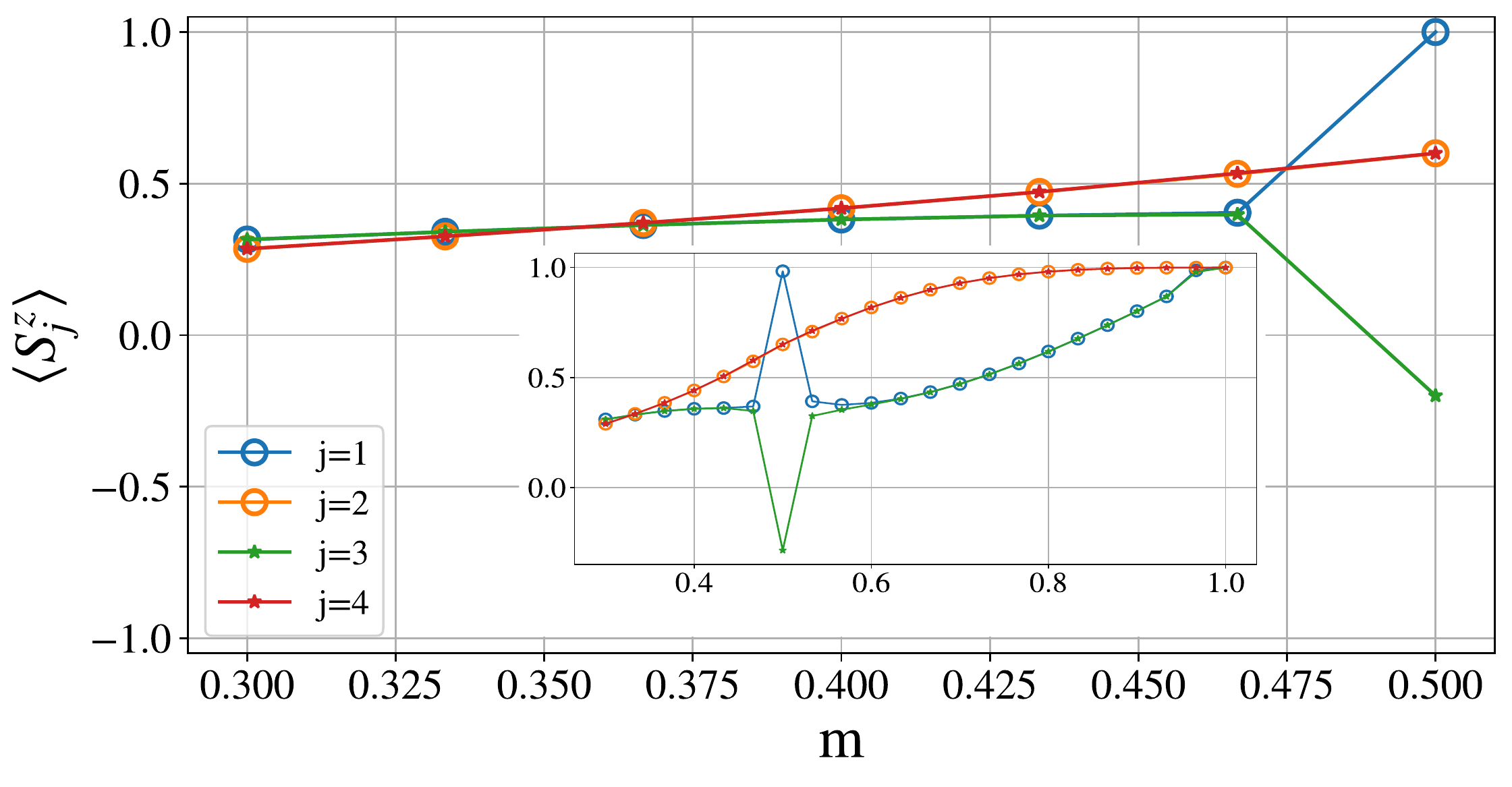}
\caption{Local normalized magnetizations in the unit cell as functions of total magnetization for $D=0$, $\alpha=\alpha_c$ and $N=60$ sites. 
The spin imbalance between the spins $\boldsymbol{S}_1$ and $\boldsymbol{S}_3$ occurs at the magnetization plateau, where $m=1/2$. There, 
the spin $\boldsymbol{S}_1$ is completly polarized, i.e, 
$ \langle S_1^z \rangle=1$ (see Fig. \ref{fig:sawtooth}).
Lower magnetization sectors are not shown for clarity. Inset: We change the frustration coupling, 
setting $\alpha=\tilde{\alpha}=\alpha_c + \frac{\alpha_c}{10}$ and the spin imbalance is still present. In both main
figure and inset $\langle S^z_2 \rangle  = \langle S^z_4 \rangle$ and the respective curves are overlapped.}
\label{fig:imbalance}
\end{figure}

We start by showing the numerical study of the spin imbalance mechanism explained in Section \ref{LEEM}. All numerical calculations 
using DMRG where made using periodic boundary conditions, unless otherwise stated. In Fig. \ref{fig:imbalance} we show the local magnetization 
$\langle S^z_j \rangle$ for each spin in the unit
cell as a function of the total normalized magnetization of the system, obtained by DMRG at $\alpha=\alpha_c$ for a $60-$sites system in the 
absence of DM interactions. The largest $m$ value showed is $m=1/2$, where the system is gapped, previous to the jump to saturation.
The breaking of the lattice translation symmetry is clearly seen only on the plateau. Although the phenomenon of exact factorization and a jump 
in the magnetization curve is exclusive to the critical value $\alpha_c$, it is important to stress that, as the path integral analysis shows, 
the phenomenon of the magnetization plateau and a spin imbalance and short range entanglement entropy on top of it is not exclusive to this critical 
value and 
happen for a finite region of the parameters. 
Indeed, in the inset of Fig. \ref{fig:imbalance} we show  again the local magnetization $\langle S^z_j \rangle$ for each spin in the unit cell as a 
function of the total normalized magnetization of the system, but we set $\alpha= \alpha_c + \frac{\alpha_c}{10} \equiv \tilde{\alpha}$ so the system does 
not have a magnetization jump to saturation and the ground state magnetization is a continuous function of $h$, but we can see that the spin imbalance 
is still present. This is to be expected due to the presence of a magnon-crystal phase (not only a fine tuned point), as the one present in the
Kagom\'e-stripe ladder\cite{acevedo2019magnon}, which also exhibits a flat band and an exact solution of localized magnons\cite{schulenburg2002macroscopic}.  

%


In the following, we present a numerical study for the spin current in the presence of DM interactions. 
To define it, we start from the  Hamiltonian on the Sawtooth lattice written as

\begin{equation}
 H=\sum_{\langle l,l' \rangle }J_{l,l'}(e^{i\theta_{l,l'}}S^{+}_{l}S^{-}_{l'}+
 e^{-i\theta_{l,l'}}S^{-}_{l}S^{+}_{l'}+\Delta_{l,l'}S^{z}_{l}S^{z}_{l'}),
 \label{eq:hamiltonian_sawtooth_l}
\end{equation}
%
%
where the angle $\theta_{l,l'}$ 
is defined by the relation $\boldsymbol{D}_{l,l'}=\check{\boldsymbol{z}} J_{l,l'}\sin{\theta_{l,l'}}$, being $\boldsymbol{D}$ the DM vector.
All throughout the work we set $\theta$ positive and  equal for all the bonds, i.e., $\theta_{l,l'}\equiv \theta>0$. The ordering in \eqref{eq:hamiltonian_sawtooth_l} 
for spin operators, which is relevant in the presence of DM interactions, is taken as in \eqref{eq:sawtooth_hamiltonian}.

\begin{figure}[t!]
\includegraphics[width=1.0\columnwidth]{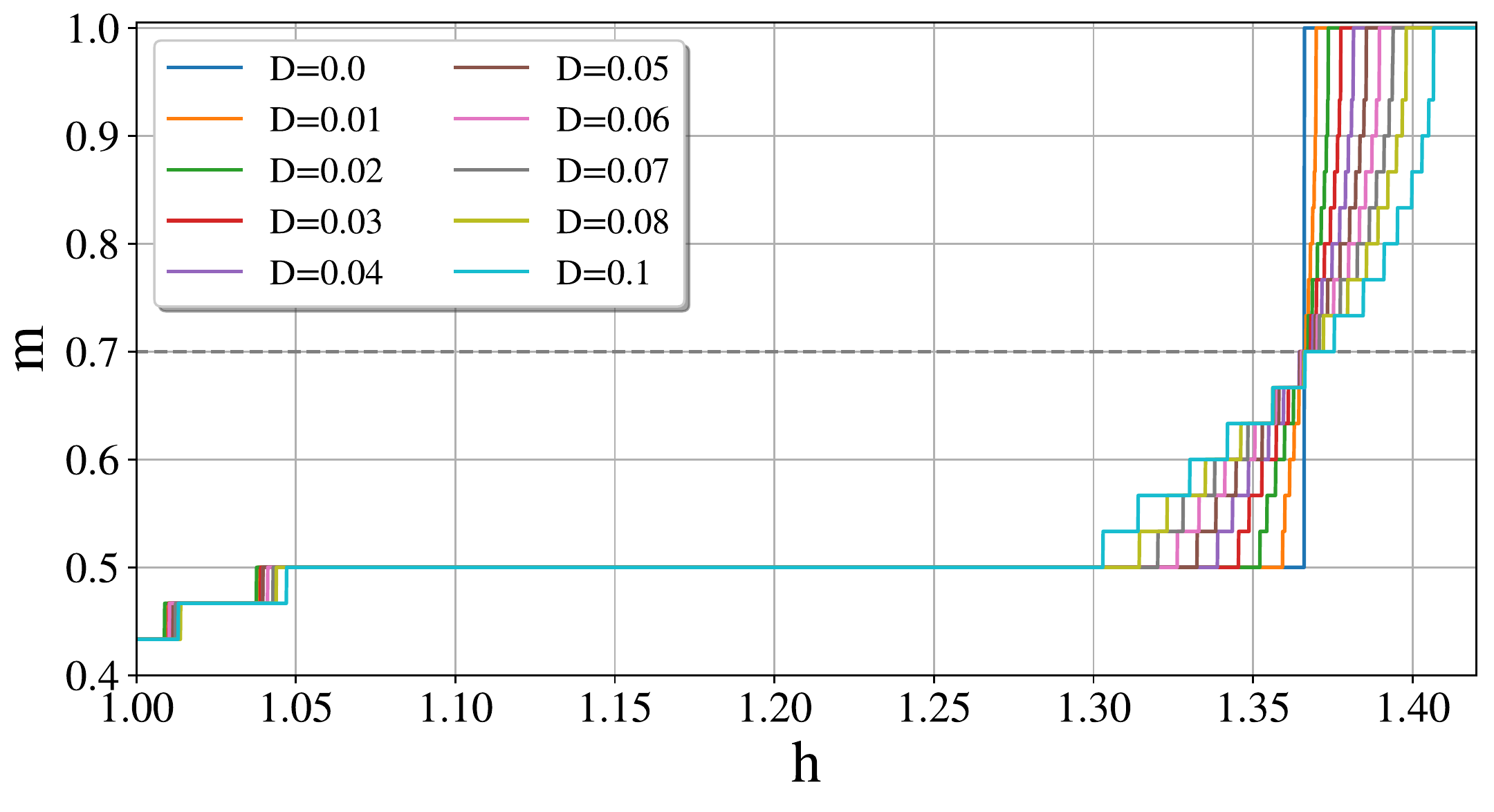}
\caption{Magnetization as a function of the applied magnetic field for different values of the Dzyaloshinskii-Moriya interaction $D$, at 
the critical coupling $\alpha_{c}$, for a $N=60$-site system.
At $D=0$ a macroscopic jump in the magnetization profile is observed as a manifestation of the magnon condensation. 
Lower magnetization sectors are not shown for clarity. }
\label{fig:mag_c}
\end{figure}

\begin{figure}[t!]
 \includegraphics[width=1.0\columnwidth]{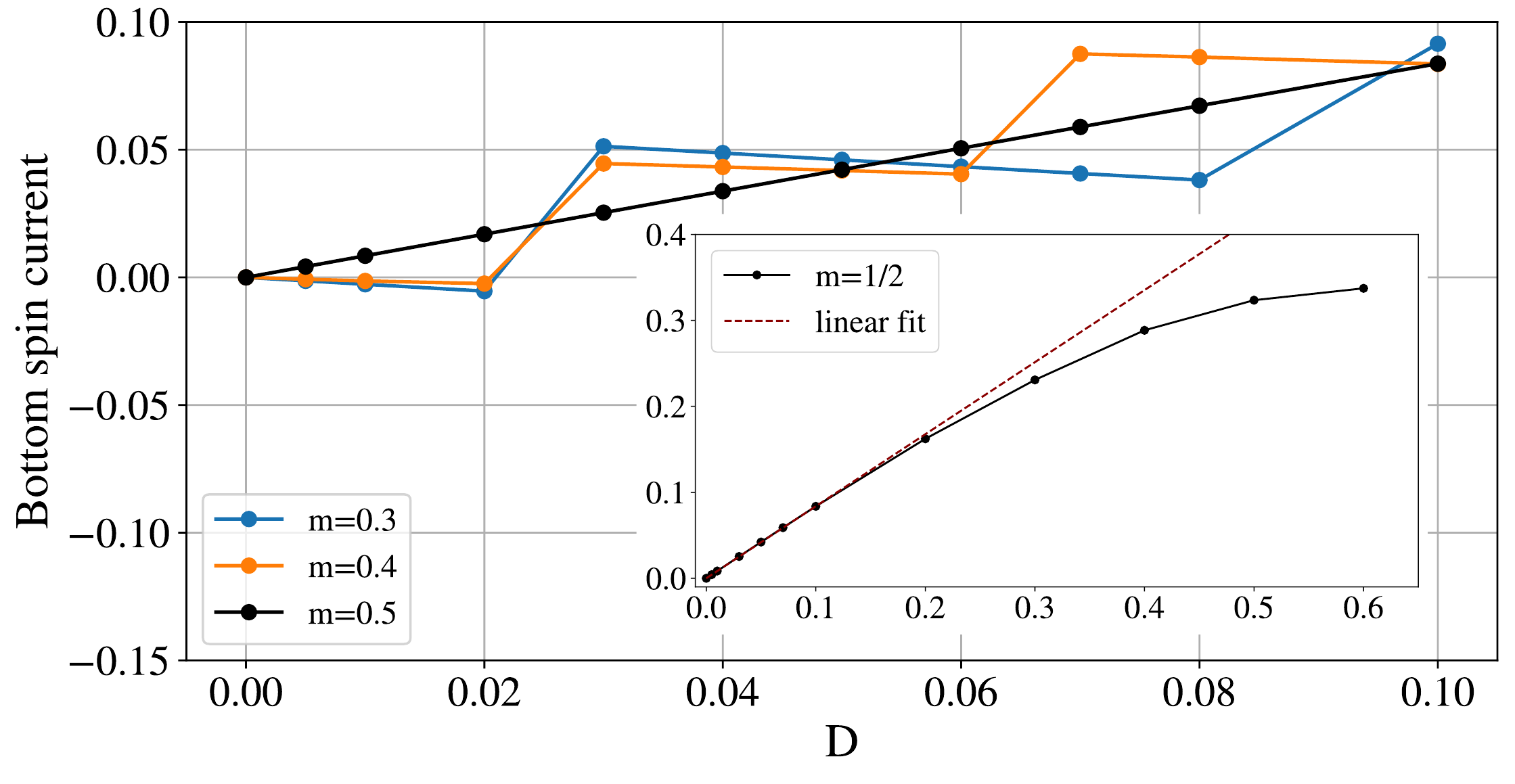}
 \caption{Spin current as a function of the Dzyaloshinskii-Moriya interaction $D$, corresponding to the critical coupling $\alpha_{c}$, and $N=60$ sites. The sawtooth profile for the current
 is a consequence of the Luttinger liquid behavior at low magnetic field. Black symbols correspond to $m=1/2$ plateau where a Meissner phase is clearly observed. Inset: Spin current
 as a function of $D$ corresponding to $m=1/2$. Notice that at high values of $D$ the system departs from the linear behavior associated to the Meissner phase.}
\label{fig:currents_low_m}
\end{figure}

\begin{figure}[t!]
\includegraphics[width=0.95\columnwidth]{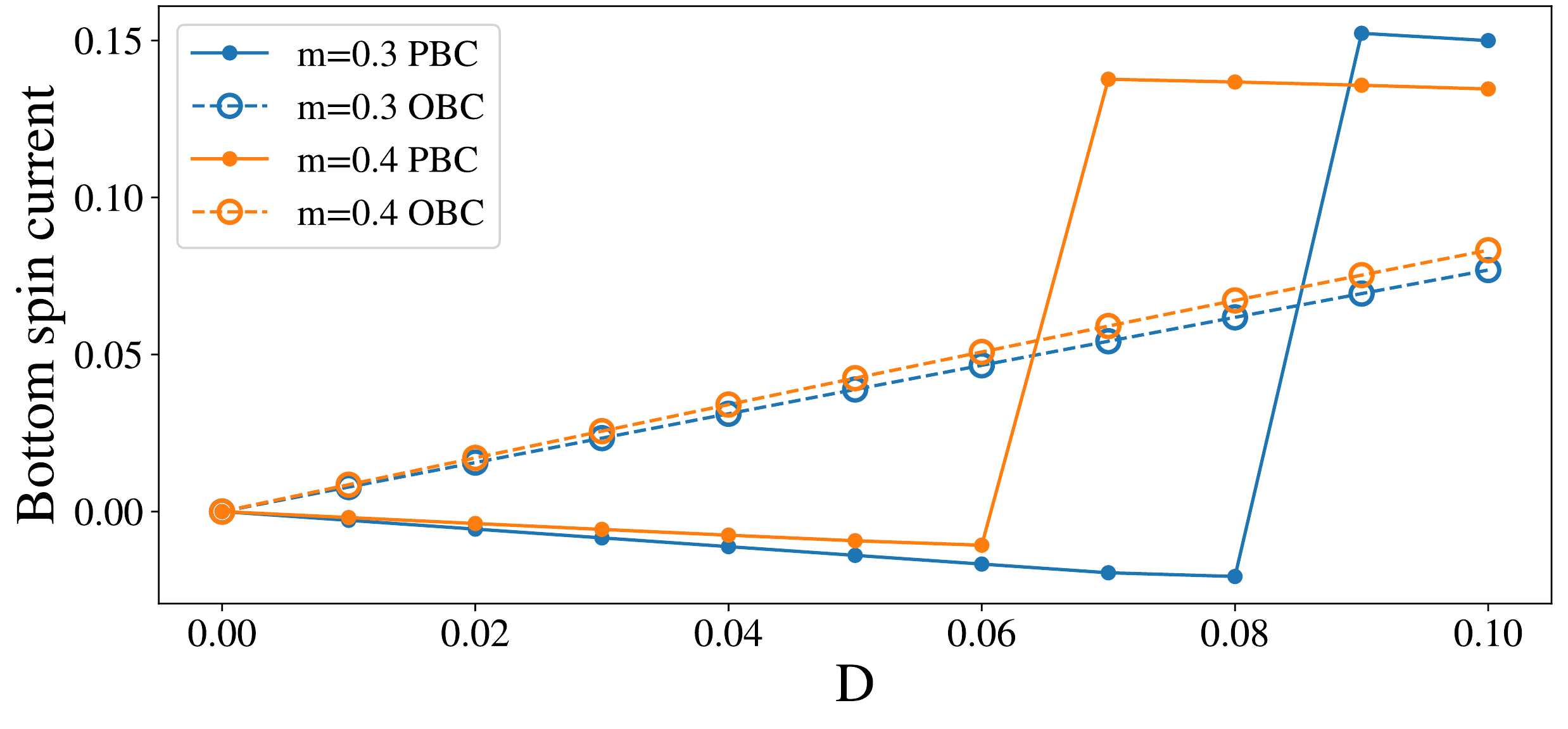}
\caption{Spin current as a function of DM interactions, at fixed magnetization, for a $20$-spin system.
In dashed lines (solid lines), the results using open (periodic) boundary conditions. The discontinuities in the spin current for $D$ in  
range $(0,1/10)$, described by the Luttinger-liquid theory, are a consequence of the periodic boundary conditions taken in the numerical calculations.}
\label{fig:OBCPBC5}
\end{figure}

\begin{figure}[t!]
\includegraphics[width=1\columnwidth]{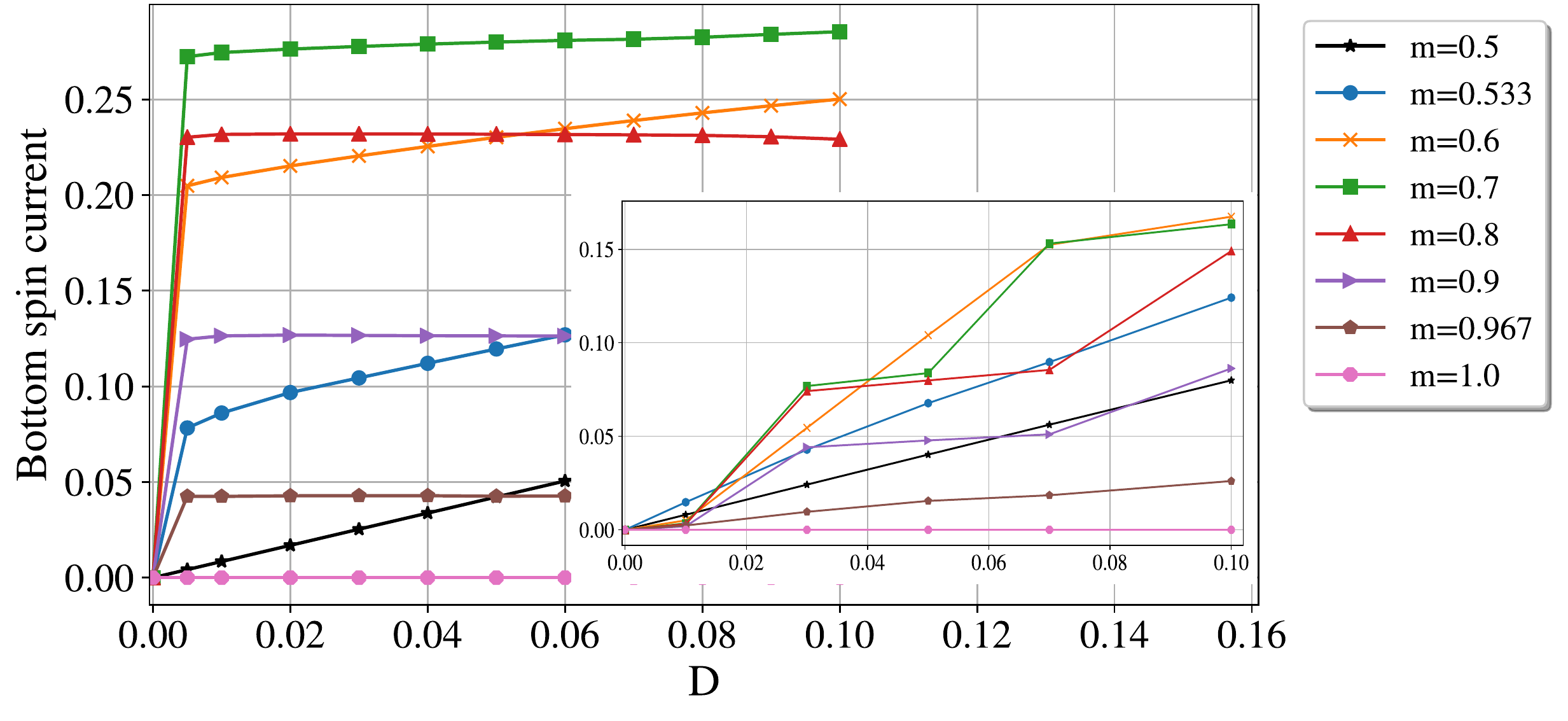}
\caption{Spin current as a function of the Dzyaloshinskii-Moriya interaction corresponding to the critical coupling $\alpha=\alpha_c$, $N=60$ sites, for different fixed magnetization sectors. At $D=0$ the current is discontinuous for $1/2<m<1$, due to the flat band. Inset: Spin current as a function of $D$, setting  $\alpha=\tilde{\alpha}=\alpha_c + \frac{\alpha_c}{10}$. The system no longer has a flat band and the spin current converges continuously to cero as $D$ tends to cero.}
\label{fig:currents_c60}
\end{figure}

We calculate numerically the ground state energy corresponding to each magnetization sector for different values of $D$ and from these
values we have built the
normalized  magnetization, $m=M/M_{sat}$, as a function of the magnetic field  for several values of $D$. The results are shown in 
Fig.  \ref{fig:mag_c}, where
a macroscopic magnetization jump from $m=1/2$ to saturation can be clearly observed for $D=0$. 
A magnetization plateau is present at $m = 1/2$. This plateau is consistent with the OYA theorem\cite{OYA} provided that the ground state unit cell
contains 4
spins, as in Fig. \ref{fig:sawtooth}. 
Although the plateau is present for all the $D$ values studied, the width of the plateau depends on this value. 
As can be seen, the dependence of the plateau edge with $D$ is greater on the right side of the plateau. 
We will see in what follows that the behavior with $D$ on both sides of the plateau is also different for spin currents.
Notably at $m=0.7$ there is virtually no dependence on the values of $D$. In the following we will see that a similar behavior occurs for the spin current, 
which can be obtained as
\begin{equation}
j_{l,l'}=i e^{i \theta_{l,l'}}\langle S^{+}_l S^{-}_{l'} \rangle  + h.c. \propto 
\langle \partial H /\partial \theta_{l,l'} \rangle
\label{eq:spin_current_def}
\end{equation}
by determining $ \langle S^{+}_l S^{-}_{l'} \rangle $  by DMRG calculations.  
The spin current is a conserved current in the sense that it satisfies a conservation law. Taking the commutator between
$S^z_j$ ($z$-magnetization density) and $H$, the corresponding result can be written as a discrete divergence of \eqref{eq:spin_current_def} 
with an extra minus sign.
In Figs. \ref{fig:currents_low_m}, \ref{fig:currents_c60}, \ref{fig:jvsh_bot} and \ref{fig:OBCPBC5} we plot the `Bottom spin current' which 
corresponds to $j_{1,3}$
in the unit cell, which is slightly larger in modulus and opposite in sign to the `Top spin current', i.e., $j_{1,2}=j_{2,3}=j_{3,4}$.

In figure \ref{fig:currents_low_m} we show the spin current obtained numerically by DMRG as a function of the DM interaction at the critical
value $\alpha_c$ for $m=0.3$, $m=0.4$ and $m=0.5$. 
As we are fixing the magnetization of the system in Figs. \ref{fig:currents_low_m} and \ref{fig:currents_c60}, it is natural to make use of a 
mapping to the bosonic system (fixing $m$ in the spin language corresponds to fixing the particle number of the bosons).


The mapping between spin and bosonic systems is a well known subject 
\cite{auerbach2012interacting,giamarchi2003quantum}.
Three common maps are the Holstein-Primakoff bosons, usually used to describe spin waves in the semiclassical limit, the Schwinger bosons, used in mean field calculations, $SU(N)$ representations and path integral descriptions, and the hard-core bosons,
which corresponds to the infinite on-site repulsion in the Bose-Hubbard Hamiltonian. The later is mainly summarized as follows:
the ladder spin operators are mapped to creation and annihilator bosonic operators, $ S^-_{j}=b^\dagger_{j}$,  the $z$-magnetization is mapped to the particle
density via $1/2 - S^z_{j} = n_{j}=b^\dagger_{j} b_{j}$,
the magnetic field $h$ is mapped to the chemical potential $\mu$, the gauge field $\boldsymbol{\theta}$ is mapped to the vector potential $\boldsymbol{A}$,
the
spin current is mapped to a particle current, the magnetization plateau corresponds to a Mott insulator phase, and the gapless phase corresponds to a 
bosonic superfluid phase.
We use the hard-core boson exact mapping to write \eqref{eq:spin_current_def} as
\begin{equation}
\begin{split}
  j_{ll'}&= i e^{i \theta_{ll'}} \langle b_l b_{l'}^\dagger \rangle + h.c. \\
\end{split}
\end{equation}
where now, the spin current is mapped to a bosonic current.
At the plateau ($m=1/2$), and for small enough values of $D$, the system remains gapped and the current becomes linear with the flux. 
This is the Meissner phase corresponding to black symbols in Fig. \ref{fig:currents_low_m}. 
In the inset, we show how for larger $D$ values the system departs from the Meissner phase because of the appearance of vortices \cite{Orignac_2016}.
Bellow the magnetic plateau ($m=0.3$ and $m=0.4$ in Fig. \ref{fig:currents_low_m}), the system remains in a Luttinger liquid phase \cite{huber2010bose}
where, under periodic boundary conditions, the Luttinger liquid 
theory can be used\cite{schulz1998} to obtain the particle current induced by a magnetic flux $\Phi$ threading the ring, giving
\begin{equation}
 j=\frac{u K }{L}( \nu -2\frac{\Phi}{\Phi_{0}}) 
 \label{eq:shulz_current}
\end{equation}
where $u$ is the spin-wave velocity, $K$ is the Luttinger parameter, $L$ is the perimeter of the ring, $\nu=N_+ - N_-$ is the difference between right moving and left moving particles, and
$\Phi_{0}=h c / e $ is the quantum of flux. 
At equilibrium, $\nu$ is chosen by the system as to minimize the energy and it can change only by integer values. 
It can be seen that the current has a periodicity $\Phi_{0}$, giving rise to a sawtooth (discontinuous) profile as a function of $\Phi$ 
which appears in the spin current discontinuities in Fig. \ref{fig:currents_low_m}.
%
%
We now discuss that numerically, this discontinuities are a direct consequence of the periodic boundary conditions taken in the DMRG calculations.
In particle language, to get effects from the gauge field, the equivalent bosonic system must have a finite $\boldsymbol{A}$ circulation over a 
close path in the system. For the sawtooth chain we have: (i) Each triangle around which we have a finite $\boldsymbol{A}$ circulation and (ii) 
the whole system with periodic boundary conditions, thought as a ring, around which we also have a finite $\boldsymbol{A}$ circulation. 
The first path makes the current in Fig. \ref{fig:currents_low_m} not periodic in $D$.
Nonetheless, the second path is much longer than the first one, so in Fig. \ref{fig:currents_low_m} for small $D$  we do not reach the first jump due to 
the cell structure.
In Fig. \ref{fig:OBCPBC5} we plotted the spin current as a function of $D$ for $m=0.3$ and $m=0.4$ for both open and periodic boundary 
conditions showing that for this range of $D$ values the discontinuities in the spin current disappear with open boundary conditions.

Let us now discuss the situation above the $m=1/2$ plateau. In figure \ref{fig:currents_c60} we show the spin current obtained numerically by DMRG as
a function of the DM interaction at the critical value $\alpha_c$ for $ m \geq 0.5 $. For $m=1/2$ the system is in the Meissner phase, shown as reference.
For $1/2<m<1$ at $D=0$ there is no spin current, but for finite D we observe a jump, which is produced by the annulation of the spin
stiffness\cite{helicity-superfluidity1973} due to the flat band. The jump is then followed by an approximately linear behavior, corresponding again to the
diamagnetic contribution as can be observed by the fact that the slope in the linear regime is simply proportional to the density of bosons. \\
At $\alpha=\alpha_c$ and $D=0$, the bosonic system has a flat band and the spin system does not accessess any of the states with $1/2<m<1$, because its 
energetically favorable for it to be either gapped ($m=1/2$) or saturated ($m=1$), depending on the applied magnetic field $h$. 
Nontheless, in Fig. \ref{fig:currents_c60} we show that, if in the bosonic system we fix the particle number to $0<n<1/4$, then the current is a 
discontinuous function of the gauge field $\theta$ at $\theta=0$.\vspace{.2 cm}\\
If we now consider the system as a spin system, and allow the magnetization to vary, the behavior can be understood from what is discussed above.
The ground state spin currents, computed by DMRG for $\alpha=\alpha_c$, are presented in Fig. \ref{fig:jvsh_bot}, where three regimes are clearly seen. 
For $h < h_{c_1}$ the system is in the Luttinger liquid phase (Fig. \ref{fig:currents_low_m}), where $m<1/2$  (Fig. \ref{fig:mag_c}). The spin current 
is a discontinuous function of $h$ because as the magnetic field is changed, the system accesses differents ground states, with different magnetizations
and possibly  a different value of $\nu$ (see Eq. \eqref{eq:shulz_current}), which is chosen by the system on \textit{each} ground state as to minimize 
its energy.
For $h_{c_1} < h < h_{c_2} $ the system is gapped, in the Meissner phase, with $m=1/2$. 
Finally, for $h_{c_2} < h < h_{sat}$ the system is in a low-stiffness phase, with $1/2<m<1$. 
In the low energy effective model from section \ref{LEEM}, the spin current \eqref{eq:spin_current_def} is $j(x) \propto \partial_x \phi $. 
 For finite $D$, at $\alpha=\alpha_c$ the renormalized spin stiffness is finite but small, and it is energetically more favorable for the system to 
 have spin currents (see Eq. \ref{eq:SKinetic_final}).
The peak arround $h=1.36$
corresponds to the maximum for the spin current at $m=0.7$ in Fig. \ref{fig:currents_c60}. At the same magnetization value all magnetization curves cross
each other in Fig. \ref{fig:mag_c}.

%


\section{flat bands in Kagome strip ladders}
\label{sec:kagome-ladder}
\begin{figure}[t!]
\includegraphics[width=0.95\columnwidth]{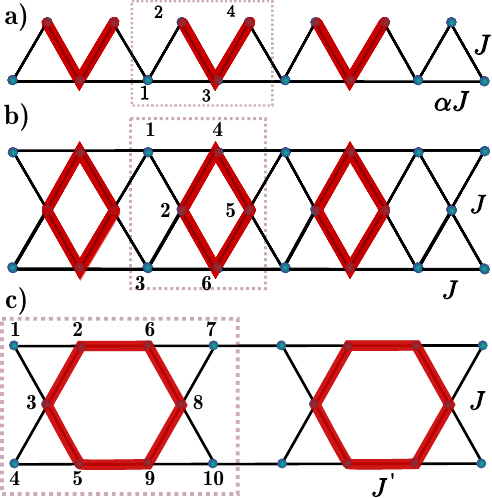}
\caption{Spin ladders that 
 for $\alpha=\alpha_c$ and $J'=J(2\Delta +1)/(\Delta+1)$ exhibit a flat band,
leading to a magnetization plateau of localized magnons and a magnetization jump to saturation. Sawtooth chain is shown here for comparison with the Kagom\'e-like ladders (b) and (c).
With dotted gray line we denote the unit cell that breaks the lattice translation symmetry. In red thick line we denote the sites in which there is a finite probability to find a magnon when each ladder has a flat band. 
}
\label{fig:flat_ladders}
\end{figure}

In Ref. \onlinecite{schulenburg2002macroscopic} there are several quantum antiferromagnets where geometrical frustration can induce a flat
band that gives place to a plateau of localized magnons and a magnetization jump to saturation.
These systems are described by a spin $1/2$ anisotropic Heisenberg Hamiltonian with first-neighbor interactions
\begin{equation}
 H= \sum_{ \langle ij \rangle } J_{ij} \bigg ( \Delta S_i^z S_j^z + \frac{1}{2} (S^+_i S^-_j +S^-_i S^+_j  )
 \bigg )
 - h S^z
\end{equation}
The work focuses in the 2D Kagom\'e lattice, but also shows that in spite of the dimensionality difference, in 1D there are three frustrated ladders that 
exhibit a flat band, here sketched in Fig. \ref{fig:flat_ladders}, where the Sawtooth is the simplest one.
It is known that for specific values of couplings shown in Fig.\ref{fig:flat_ladders} and chosen in this section, the magnetization jumps to saturation 
are of magnitude $\delta m=1/3$ and $1/5$ for ladders $b)$ and $c)$ in Fig. \ref{fig:flat_ladders}, respectively.


Following the steps presented in section \ref{LEEM} we study the presence of possible delocalized modes by studying the mass matrix for 
 the two ladders that present a Kagom\'e-like
structure  in Fig \ref{fig:flat_ladders}-b) and  \ref{fig:flat_ladders}-c).
\begin{figure}[t!]
\includegraphics[width=0.95\columnwidth]{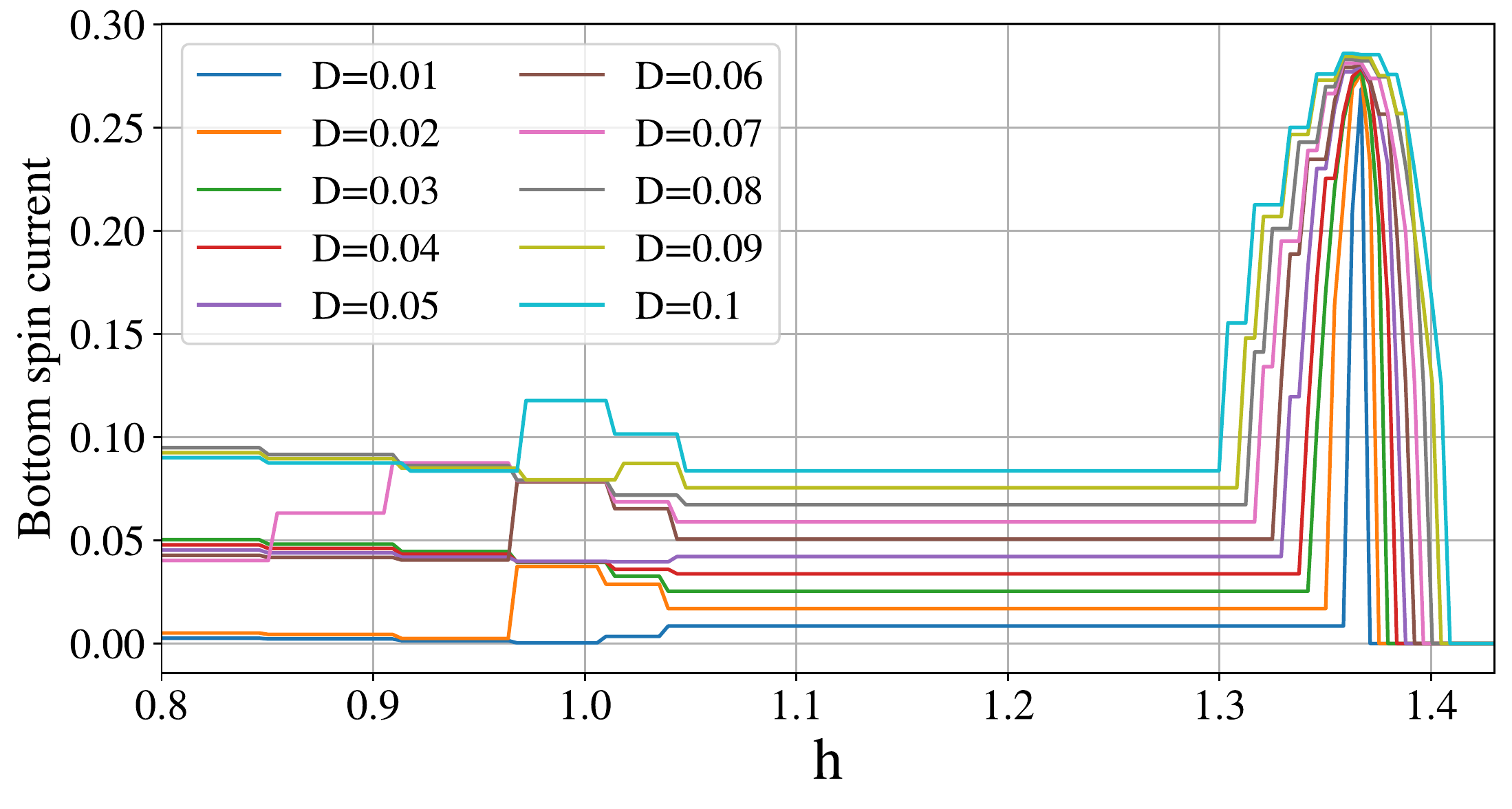}
\caption{Ground state spin current as a function of the applied magnetic field for diferent $D$ values, for the critical coupling $\alpha=\alpha_c$ 
and $N=60$.
As $h$ varies, the system accesses different magnetization sectors as we show in Fig. \ref{fig:mag_c}. Here, for the ground state spin current there are
three regimes. First, the low-field regime ($m<1/2$), where the system is in the Luttinger-liquid phase and the ground state spin current is a discontinuous
function of $D$. Increasing the applied magnetic field, the system enters in the Meissner phase, it is gapped, and has $m=1/2$. Finally, at high field 
the system is in the low-stiffness phase, with higher continuous spin currents and $1/2<m<1$. }
\label{fig:jvsh_bot}
\end{figure}

Let us consider first the Kagom\'e-like ladder represented in Fig. \ref{fig:flat_ladders}-b).  By symmetry, we can take the following anzats for the
classical polar angles: 
$\theta_1^{(b)}=\theta_3^{(b)}=\theta_4^{(b)}=\theta_6^{(b)} \equiv \theta_A^{(b)}$, $\theta_2^{(b)}=\theta_5^{(b)}\equiv \theta_B^{(b)}$. 
For the classical azimuthal angles we take  $\phi_2^{(b)}=\phi_5^{(b)}$ and $\phi_l^{(b)}-\phi_2^{(b)}\equiv\phi^{(b)}$ for $l=1,3,4,6$. 
%
The Kagom\'e-like ladder in Fig \ref{fig:flat_ladders}-b)  presents a magnetization plateau and localized magnons at $m=2/3$. 
The classical ground state corresponding to this magnetization sector has again $\phi^{(b)}=\pi$. 
Adding quantum fluctuations as in \eqref{eq:fluctuations}, and taking the continuum limit, we construct the mass matrix, $M_\phi^{(b)}$, for 
the $\phi_l^{(b)}(x)$ \textit{fields}. As before, this symmetric mass matrix can be diagonalized by a unitary transformation,
\begin{equation}
P^{(b)}=\left(
\begin{array}{cccccc}
 \frac{1}{\sqrt{6}} & \frac{1}{\sqrt{6}} & \frac{1}{\sqrt{6}} & \frac{1}{\sqrt{6}} & \frac{1}{\sqrt{6}} & \frac{1}{\sqrt{6}} \\
 0 & 0 & -\frac{1}{\sqrt{2}} & 0 & 0 & \frac{1}{\sqrt{2}} \\
 -\frac{1}{\sqrt{2}} & 0 & 0 & \frac{1}{\sqrt{2}} & 0 & 0 \\
 -\frac{1}{2} & 0 & \frac{1}{2} & -\frac{1}{2} & 0 & \frac{1}{2} \\
 0 & -\frac{1}{\sqrt{2}} & 0 & 0 & \frac{1}{\sqrt{2}} & 0 \\
 \frac{1}{2 \sqrt{3}} & -\frac{1}{\sqrt{3}} & \frac{1}{2 \sqrt{3}} & \frac{1}{2 \sqrt{3}} & -\frac{1}{\sqrt{3}} & \frac{1}{2 \sqrt{3}} \\
\end{array}
\right),
\label{eq:P_b}
\end{equation}
giving $$ M_{\phi}^{(b)'}=P M_\phi^{(b)} P^t=\text{Diag}(0,m_2^{(b)},m_3^{(b)},m_4^{(b)},m_5^{(b)},m_6^{(b)}),$$ where

\begin{equation}
 \begin{cases}
  m_2^{(b)}=m_3^{(b)}=2(\gamma^{(b)}-2\beta^{(b)}) \\
  m_4^{(b)}=2 \gamma^{(b)} \\
  m_5^{(b)}=4 \gamma^{(b)} \\
  m_6^{(b)}=6 \gamma^{(b)}
 \end{cases}
 \label{eq:M_b}
\end{equation}
with $\beta^{(b)}=\sin^2{\theta_B^{(b)}}$ and    $\gamma^{(b)}=\sin{\theta_A^{(b)}} \sin{\theta_B^{(b)}}$.
The  transformed fields are 
$\varphi^{(b)}_l= \sum_{l'} P^{(b)}_{l,l'} \phi^{(b)}_{l'}$.
Equations \eqref{eq:P_b} and \eqref{eq:M_b} mean that the combinations $\varphi^{(b)}_j$ are gapped, for $j=4,5,6$, and in the low energy limit they vanish. 
The gapless fields 
$\varphi_1^{(b)}$ represents, as before, the Goldstone mode associated to the $U(1)$ rotation symmetry of the system.
The mass $m_2^{(b)}$ can vanish for classical angles $\theta_A^{(b)}$ and $\theta_B^{(b)}$ such that $\gamma^{(b)}=2\beta^{(b)}$. 
Then, the fields $\varphi_2^{(b)}=(\phi_6^{(b)}-\phi_3^{(b)})/\sqrt{2}$ and $\varphi_3^{(b)}=(\phi_4^{(b)} - \phi_1^{(b)})/\sqrt{2}$ get delocalized simultaneously.
This leads again to a spin imbalance mechanism that describes the localized excitations as in the sawtooth chain.   

For the Kagom\'e-like structure represented in Fig \ref{fig:flat_ladders}-c) we take a classical ground state with a 10-site unit cell. 
The polar angles corresponding to the classical ground state are $\theta_l^{(c)}\equiv \theta_A^{(c)}$ for $l\, \epsilon\, \{1,2,4,5,6,7,9,10\}$, 
and $\theta_3^{(c)}=\theta_8^{(c)}\equiv \theta_B^{(c)}$.
The azimuthal angles are $\phi_l^{(c)}\equiv \phi^{(c)}$ for $l \, \epsilon \, \{1,2,4,5,8 \} $ and 
$\phi_{l'}^{(c)} - \phi^{(c)}= \pi $ for $l' \, \epsilon \, \{3,6,7,9,10 \} $.
Following the same steps as in the previous cases, we construct the mass matrix for the $\phi_l^{(c)}(x)$ fields. 
It can be seen that the combinations $\phi_1^{(c)}-\phi_2^{(c)}$, $\phi_4^{(c)}-\phi_5^{(c)}$, $\phi_7^{(c)}-\phi_6^{(c)}$, $\phi_{10}^{(c)}-\phi_9^{(c)}$ 
can simultaneously become massless for a particular value of $\theta_A^{(c)}$ and $\theta_B^{(c)}$, in the low energy limit. 
The delocalization of this modes typically lead
to a region in parameter space with spin imbalance and short range entanglement, and more precisely a magnon crystal phase, among which the exact localized magnons state is a particular point. 

The delocalized modes in both Kagom\'e-like ladders are described as in the sawtooth ladder, due to the shared flat band presence. Furthermore, the flat band means a vanishing stiffness. Then, the spin currents in these Kagom\'e-like ladders must also be discontinous at $D=0$ for fixed magnetization sectors previous to saturation, and a low-stiffness phase (as in Fig. \ref{fig:jvsh_bot}) is to be expected in both these ladders. 
Finally we must remark that the path integral formalism based on coherent states it is \textit{not} restricted to one dimensional systems, 
as shown in \onlinecite{TTH,PI2}, and neither is the case for delocalized modes due to vanishing masses in the theory. 
Then, we expect that the flat band in the Kagom\'e lattice also gives place to discontinuous spin currents at $D=0$ and a low-stiffness phase, 
but such study exceeds the scope of this work.


\section{Conclutions}
\label{sec:conclutions}
We have studied the sawtooth chain in the case where frustration induces a flat band, and how DM interactions affect this particular system.
Using a semiclassical field theory approach we have described the localized excitations through a spin imbalance mechanism. 
A central aspect of this mechanism is the presence of delocalized angular modes whose presence can be detected by diagonalizing the mass matrix. 
%
This matrix can also be evaluated for different spin systems (even in higher dimensions) where one also expect flat magnon dispersion. In particular we 
have done it for two Kagom\'e-like ladders that present flat bands, showing the relationship between the delocalized modes there, and the corresponding 
localized magnons excitations. 
We have studied numerically the spin current on the sawtooth chain introducing antisymmetric interactions, finding three 
different regimes. At small values of the applied magnetic field, the system is in a Luttinger liquid phase. In this phase, the spin current shows a step
behavior as a 
function of parameter $D$. This behavior can be clearly understood through Luttinger-liquid theory. At intermediate values of the magnetic field, the 
magnetization 
curve presents a plateau at $m=1/2$, being $m$ the total normalized magnetization. In this plateau, the ground state is a gapped magnon crystal in the
absence of DM interactions. For finite $D$ the spin current is constant on the plateau and proportional 
to the DM parameter as a consequence of the diamagnetic term of the spin current. This phase is labeled as the Meissner phase. Finally, at the high magnetic
field, the
dependence with $D$ is notorious. At $D=0$ there is a jump in the magnetization curve. The magnetic sectors between $m=1/2$ and $m=1$ are skipped. Then for $D=0$ there is no high magnetic field phase. 
However, for a finite value of $D$, the magnetization curve is smooth as the field $h$ varies. The ground state spin current is in this case also smooth, and has a peak at roughly $h/J=1.36$. In terms of the effective field theory, the system has no delocalized modes and has a finite but small spin stiffness.
If instead, the magnetization is kept fixed at $1/2<m<1$, then the spin current presents a jump at $D=0$, as a consequence of the flat band.

From a more general stand point, we have used as a laboratory the Sawtoooth ladder to develop general arguments that allow to study two interesting 
properties of systems with a flat band. The first is the fact that, in the parameter space, a whole magnon crystal phase is present around the critical
point corresponding to the flat band. In this magnon crystal phase the system shows a spontaneous translation symmetry breaking and short range entanglement, 
an issue that can reveal very interesting in the very active subject of scar states \cite{Turner}. The second property concerns the singular
behaviour of the currents with respect the presence of a small DM interaction for spin systems or flux for hard core bosons. Our arguments are
general enough to confidently predict the very same properties for the other systems that house a flat band, like the kagome strip ladders or even the 2-D
kagome lattice.


\section*{Acknowledgments}
We would like to thank the "Laboratoire International Associ\'e" LIA LICOQ and the LABEX NEXT for support.
C. A. Lamas is also supported by ANPCyT (PICT 2013-0009)

\vspace{1cm}

\bibliography{referencias-sawtooth}


\end{document}